\documentstyle[twoside,fleqn,epsf,espcrc2]{article}
 
\newcommand{\AmS}{{\protect\the\textfont2
  A\kern-.1667em\lower.5ex\hbox{M}\kern-.125emS}}
 
\title{B and D semileptonic decays to light mesons  \hfill\normalsize FERMILAB-CONF-99/241-T}
\author{
S.M.~Ryan$^a$ \thanks{talk presented by S.~Ryan.}, A.X.~El-Khadra$^b$, A.S.~Kronfeld$^a$, P.B.~Mackenzie$^a$ and J.N.~Simone$^a$
\\[5mm]
$^a$ {\small Theoretical Physics Department, Fermilab, P.O. Box 500, Batavia, Il 60510, U.S.A.}\\
$^b$ {\small Loomis Laboratory of Physics, 1110 W. Green Street, Urbana, Il 61801-3080, U.S.A.} 
}
\begin{document}  
\begin{abstract}
We present an update of our calculation of the form factors and partial widths of semileptonic heavy-to-light meson decays, in the quenched approximation. Our complete data set has now been analysed and our results include chiral extrapolations and a study of the lattice spacing dependence. 
\end{abstract}
\maketitle
\section{Introduction}
We report on the progress made in our study of $B$ and $D$ meson semileptonic decays. A description of the analysis and some more preliminary results are in Refs.~\cite{me_lat98,jim_lat98}. 
The CKM matrix element $|V_{ub}|$ plays an important r\^ole in over-constraining the unitarity triangle but is only determined to $\sim 20\%$.
The advent of $B$-factories will reduce the experimental error in exclusive decays considerably, which must be matched by more precise theoretical determinations of the nonperturbative contribution. With the increase in data, experiments will be able to 
study the $q^2$ distribution in $B$ and $D$ semileptonic decays, as shown by the CLEO collaboration, which recently presented a new analysis of $B\rightarrow\rho l\nu$~\cite{cleo_B2rho}. 
Thus, lattice and experimental data could be combined in a range 
of $q^2$s where both are reliable, making the model-dependent 
extrapolation to $q^2 =0$, traditionally done in lattice analyses, redundant. 

A summary of our work is as follows. We have results at three lattice spacings ($\beta =5.7,5.9$ and $6.1$) with heavy quarks at the $b$ and $c$ quark masses and a light quark (daughter and spectator) at the strange quark mass. 
This allows us to study the lattice spacing dependence of the matrix elements, $\langle s\bar{s}(p)|V_\mu |B_s(0), D_s(0)\rangle$ and experience leads us to 
believe the $a$-dependence does not change after chiral extrapolation, 
see Ref.~\cite{fBpaper}. We have additional light quarks at $\beta =5.9$ and 
$5.7$ allowing chiral extrapolations on these lattices.
The quark fields are rotated but the perturbative matching coefficients we use are those of light quarks; the full mass-dependent calculation is underway. 
The final results are at $\beta = 5.9$, after chiral extrapolation.

The matrix elements are extracted from three-point correlation functions in which the heavy meson is at rest and the light meson has momentum of 
(0,0,0), (1,0,0), (1,1,0), (1,1,1) and (2,0,0), in units of $2\pi /aL$. 
In an approach different to other groups~\cite{ukqcd,ape,jlqcd} we study the $a$-dependence and perform the chiral extrapolations in terms of the matrix elements and not the form factors. 

For $B$ and $D$ decays the form factors are determined from matrix elements in the usual way,
\begin{eqnarray}
\langle\pi(p)|V_\mu |B(p^\prime )\rangle\hspace{-0.3cm}&=&\hspace{-0.3cm} f^+(q^2)\left [ p^\prime+p-\frac{m_B^2-m_\pi^2}{q^2}q\right ]^\mu \nonumber\\
&+& f^0(q^2)\frac{m_B^2-m_\pi^2}{q^2}q^\mu 
\end{eqnarray}
and thence we determine the partial widths
\begin{equation}
\frac{d\Gamma}{d|\vec{p}_\pi|} =\frac{2m_{B,D}G_F^2|V_{ub}^2|}{24\pi^3}\frac{ |\vec{p}_\pi |^4}{E_\pi}\left | f^+(q^2)\right |^2.~\label{rate_eqn}
\end{equation}
\section{$B\rightarrow\pi l\nu$}
\begin{figure}[h]
\vspace{-1.0cm}
\begin{center}
\leavevmode  
\epsfxsize=2.9in
\epsfysize=2.6in
\hspace{-0.5cm}\epsfbox{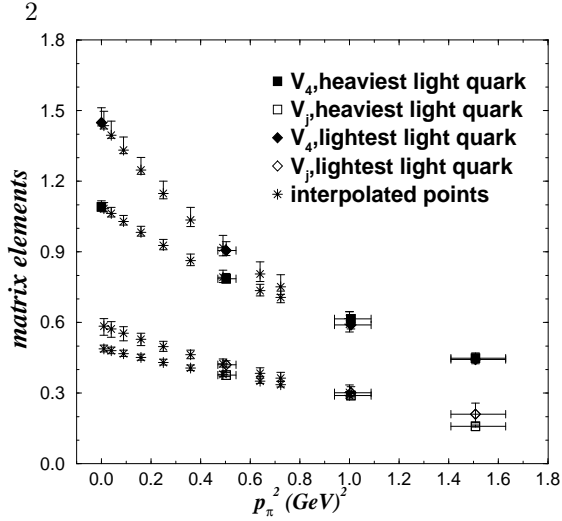}
\end{center}
\vspace{-0.58in}
\caption{Comparing matrix elements at $\beta=5.9$, with the heavy quark at the $b$ mass.}
\label{fig_mxinterp}
\vspace{-.8cm}
\end{figure}

We interpolate the spatial and temporal matrix elements, extracted from fits to the three-point functions, to fixed physical momenta in the range \{0.1,$\ldots$ ,1.0\} GeV as shown in 
Figure~\ref{fig_mxinterp}. 
A significant systematic error is evident in the interpolation of the matrix elements. 
The temporal component, $V_4$, is defined at zero momentum so all physical momenta are obtained by interpolation. In contrast, the spatial matrix element is not ... the first point is $p_{lat}(1,0,0)$ $\approx 0.7$ GeV. All momenta below this are obtained by extrapolation (see Fig.~\ref{fig_mxinterp}). 
Thus for lighter quarks the temporal component captures the effect of the nearby $B^\ast$ pole at $p=0$ and rises rapidly, whereas the spatial component misses this effect. 
Now one has a choice: introduce a model, eg. pole dominance, to reproduce the 
behaviour of the matrix elements in the vicinity of the $B^\ast$ pole or impose a cut in momentum, below which the extrapolation of the spatial matrix element is considered unreliable. 
We wish to avoid any model dependence so we make a cut in momenta at $p=0.4$ GeV.

\begin{figure}[h]
\vspace{-1.0cm}
\begin{center}
\leavevmode  
\epsfxsize=2.9in
\epsfysize=2.6in
\hspace{-0.5cm}\epsfbox{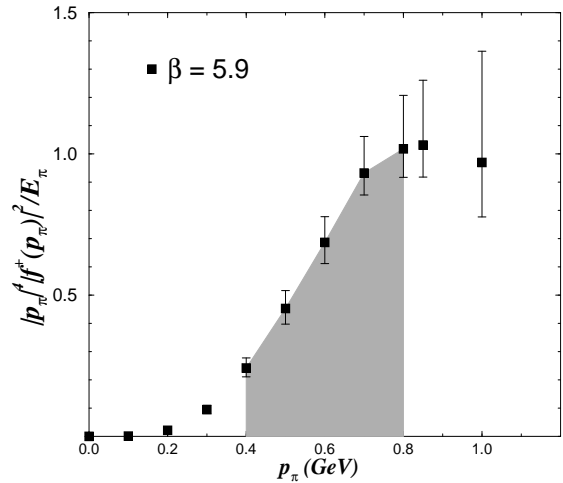}
\end{center}
\vspace{-0.58in}
\caption{Differential decay rate at $\beta =5.9$ and partial width}
\vspace{-0.8cm}
\label{rate_final}
\end{figure}

At $\beta=5.7$ and $5.9$ the matrix elements are extrapolated to the chiral limit. The data at $\beta=5.9$ are discussed here (the findings are similar at $\beta=5.7$). 
Three functional forms were compared in the chiral extrapolations: linear, quadratic and including a term $\propto\sqrt{m_q}$. For $0.4\leq p\leq 0.8$ GeV the best fit to the data was the quadratic form. At $p>0.8$ GeV all three functions resulted in unreliable fits, with bad $\chi^2/d.o.f.$, due to large cutoff effects and an increasingly noisier signal. Therefore the range of momenta we consider is $0.4\leq p \leq 0.8$ GeV.
The error due to extrapolation is estimated by considering the spread in results from the three possible fit forms, in this range.  
\begin{figure}[h]
\vspace{-1.0cm}
\begin{center}
\leavevmode  
\epsfxsize=2.9in
\epsfysize=2.6in
\hspace{-0.5cm}\epsfbox{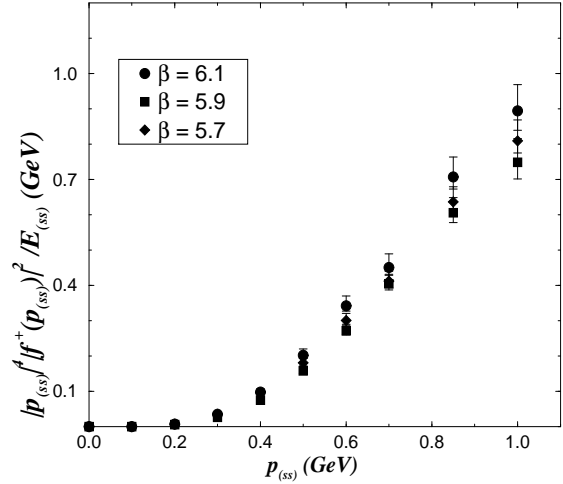}
\end{center}
\vspace{-0.58in}
\caption{Differential decay rate at 3 lattice spacings, with strange light quarks.}
\vspace{-0.8cm}
\label{rate_adep}
\end{figure}  
Figure~\ref{rate_adep} shows the $a$-dependence of our results (with strange 
light quarks) is very mild. The matrix elements show a similarly mild 
$a$-dependence~\cite{me_lat98}.
The scale to determine the physical momenta is set from the spin-averaged 1P-1S splitting in Charmonium. The quenched approximation introduces a dependence on the quantity used to set the scale and this is often used to estimate quenching effects. We repeated the procedure with $a^{-1}(f_\pi )$ and found only a small effect which is included in our error estimates, with the caveat that it is almost certainly an underestimate of quenching. 
The partial width of $B\rightarrow\pi l\nu$, for $0.4\leq p\leq 0.8$, is shown as the shaded region in Figure~\ref{rate_final}. This width and the statistical error is: $2m_B\int_{p=0.4}^{p=0.8}|p_\pi |^4f^+(p_\pi)^2/E_\pi = 12.17(11)$. 
\section{$D\rightarrow\pi l\nu /D\rightarrow Kl\nu$}
\begin{figure}[h]
\vspace{-1.0cm}
\begin{center}
\leavevmode  
\epsfxsize=2.9in
\epsfysize=2.6in
\hspace{-0.5cm}\epsfbox{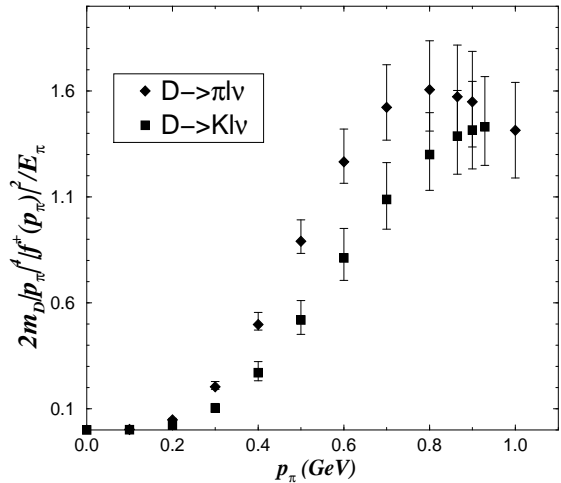}
\end{center}
\vspace{-0.58in}
\caption{A comparison of rates for D decays}
\vspace{-0.8cm}
\label{D2pi_on_D2K}
\end{figure}

\begin{figure}[h]
\vspace{-1.0cm}
\begin{center}
\leavevmode  
\epsfxsize=2.9in
\epsfysize=2.6in
\hspace{-0.5cm}\epsfbox{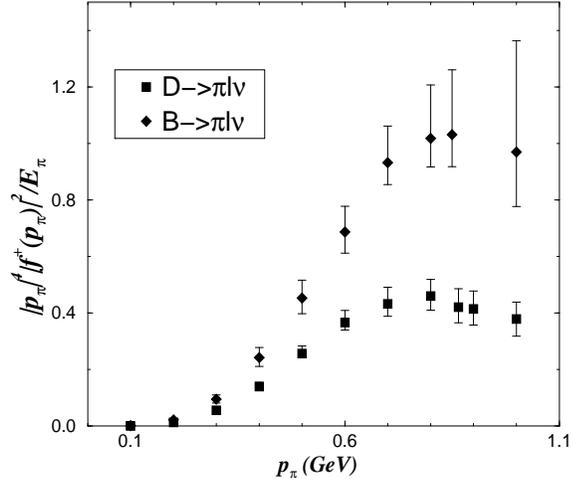}
\end{center}
\vspace{-0.58in}
\caption{The rates for B and D decays}
\vspace{-0.8cm}
\label{B2pi_on_D2pi}
\end{figure}

In Ref.~\cite{jim_lat98} it was suggested that calculating the ratio of partial widths $D\rightarrow\pi l\nu /D\rightarrow Kl\nu$ is a nice way to reduce the uncertainty on $|V_{cd}|/|V_{cs}|$, from its current $\sim 17\%$. The Focus Collaboration at Fermilab expects to have of ${\cal O}(10^6)$ fully reconstructed $D$ decays so the experimental error will be considerably reduced. 
By calculating a ratio of rates it is expected that much of the theoretical uncertainties will cancel. In particular, the renormalisation factors cancel, eliminating the perturbative uncertainty.
The analysis of $D$ decays proceeds as described for $B$ decays with further details in Ref.~\cite{jim_lat98}. 
Here, we report on the update since last year. The chiral extrapolations at $\beta=5.9$ have been done for the pion and kaon final states so we can now calculate the ratio $D\rightarrow\pi l\nu /D\rightarrow Kl\nu$, as shown in Figure~\ref{D2pi_on_D2K}. 
The ratio of partial widths in the range $0.2\leq p_\pi\leq 0.7$ is 1.61(19), where the error is statistical. 
We also calculate the ratio $B\rightarrow\pi l\nu /D\rightarrow\pi l\nu$, shown in Figure~\ref{B2pi_on_D2pi}. With the expected experimental precision in D decays this ratio has the advantage that many uncertainties are reduced and therefore may prove an interesting avenue for a determination of $|V_{ub}|$. 
\section{Conclusions}
In conclusion we present a preliminary summary of the systematic errors 
contributing to the theoretical error in $|V_{ub}|$ and $|V_{cd}|/|V_{cs}|$.
The uncertainty due to the perturbative matching is not yet included. 
Adding in quadrature gives an error of $\sim 10\%$ on $|V_{ub}|$ and 
$\sim 13\%$ on $|V_{cd}|/|V_{cs}|$ and 
we expect to improve upon this in the final result. 
\begin{tabular}{c|c|c}
	             &$|V_{ub}|$ &$\frac{|V_{cd}|}{|V_{cs}|}$ \\ 
\hline
Statistics           &4\%    & 5\% \\
$\chi$-extrapolation, p-interpolation&8\%    & 10\%  \\
$a$-dependence	     &5\%	   & 5\%   \\
Determining $a^{-1}$&3\%& 3\% \\
Fits,excited state contamination &2\%	   & 2\%	 \\
$m_Q$-dependence     &1\%    & 1\%  \\
\hline
\end{tabular}
\begin{flushleft}{\bf Acknowledgements}\end{flushleft}
Fermilab is operated by Universities Research Association, Inc. for the U.S. Department of Energy.

\end{document}